# Comprehensive Characterization of Terahertz Generation with the Organic Crystal BNA

Isaac C. Tangen,[1,†] Gabriel A. Valdivia-Berroeta, [1,†] Larry K. Heki, [1,†] Zachary B. Zaccardi,[1] Erika W. Jackson, [1] Charles B. Bahr, [1] David J. Michaelis, [1,a] and Jeremy A. Johnson[1,a]

**Affiliations**

[1] Department of Chemistry and Biochemistry, Brigham Young University, Provo, UT, 84602, USA.
[a] Authors to whom correspondence should be addressed: dmichaelis@chem.byu.edu, jjohnson@chem.byu.edu

**Abstract**

We characterize the terahertz (THz) generation of N-benzyl-2-methyl-4-nitroaniline (BNA), with crystals ranging in thickness from 123-700 μm. We compare excitation using 800-nm and 1250 to 1500-nm wavelengths. Pumping BNA with 800-nm wavelengths and longer near-infrared wavelengths results in a broad spectrum, producing out to 6 THz using a 100-fs pump, provided the BNA crystal is thin enough. ~200 μm or thinner crystals are required to produce a broad spectrum with an 800-nm pump, whereas ~300 μm thick crystals are optimal for broadband THz generation using the longer wavelengths. We report the favorable THz generation and optical characteristics of our BNA crystals that make them attractive for broadband, high-field THz generation, and we also find significant differences to BNA results reported in other works.

The down-conversion of ultrafast near-infrared (NIR) laser pulses via optical rectification has become a standard method to generate broadband terahertz (THz) pulses. Because optical rectification involves interacting pump and generated THz waves of very different frequencies inside a nonlinear optical (NLO) crystal, phase-matching can be a challenge. The phase-matching condition is fulfilled when the group index of the input light matches the phase index of the output THz wavelengths[1-3]. With the optical properties of a particular NLO crystal, we can identify crystal thicknesses and pump wavelengths that lead to optimal THz generation.

To date, organic NLO crystals like DAST[4], DSTMS[5], and OH1[3], and more recent developments, like HMQ-TMS[6] and EHPSI-4NBS[7], have produced the strongest THz fields due to very good phase-matching with 1200-1600 nm NIR pump wavelengths. The NLO organic salt crystals DAST and DSTMS show ~10 MV/cm generated THz fields when pumped with near-infrared light[4]. Nevertheless, both exhibit a detrimental dip in THz generation dip at 1 THz, limiting the applicability of these crystals as truly broadband THz sources. Complimentary to DAST and DSTMS, the hydrogen-bonded OH1 crystal displays a strong generation profile from 0 to 3 THz[3], which is similar to tilted-pulse-front THz generation with $LiNbO_3$[8-10]. Several studies have utilized these crystals to access either lower (<3 THz) or higher (>3 THz) frequencies and resonantly excite specific modes[11-15].

Crystalline N-benzyl-2-methyl-4-nitroaniline (BNA) is another interesting molecular crystal because it features weaker phonon absorptions in the THz region (compared to DAST, DSTMS, and OH1). Initially designed to improve the crystallization behavior of NLO MNA crystals, BNA has been tested for THz generation via difference-frequency generation[16-18] and optical rectification[19, 20]. THz pulses with spectral content from 0 to 3 THz were obtained via optical rectification using an 800-nm pump wavelength; however, ZnTe was used as an electro-optic (EO) detection crystal, which may have unnecessarily limited the detection bandwidth[19]. In a recent study, BNA was pumped with 1150–1550 nm, 35 fs pulses to produce THz with spectral content from 0 to 7.5 THz[20]. However, a significant dip in the THz generation spectrum was observed at ~2 THz at all the tested thicknesses from 200 μm to 2.9 mm.

In this letter, we pump BNA with 800-nm and 1250-1500-nm pump pulses and report the THz output of crystal thicknesses ranging from 123 to 700 μm. Previous reports have shown differences in BNA crystals produced with different growth methods[21], and we show that in our crystals, the previously reported dip in THz generation at 2 THz is greatly reduced. This results in a broad, smooth THz spectrum from 0 to 6 THz. We also report improved broadband THz generation from 0 to 5 THz when thin BNA crystals are pumped with 800-nm light. This greatly extends the applicability of BNA with Ti:Sapphire laser systems.

We also determine BNA absorption coefficients and refractive and group indices in the THz and NIR region, revealing important differences with previously reported values[22, 23]. With BNA rising in prominence as a useful THz generator, we believe it's useful to report that differences in crystal characteristics may arise due to differences in processing methods.

BNA crystals with thicknesses ranging from 123 to 700 μm were obtained in collaboration with THz Innovations, LLC. To test THz generation, a 1-kHz repetition-rate, ~100-fs 800-nm Ti:Sapphire laser was used to either directly pump the BNA crystals or to generate IR pump wavelengths (1250 to 1500 nm) output from an optical parametric amplifier (OPA). The BNA crystals were oriented such that the optic axis of each crystal was parallel to the vertical polarization of the pump beam. After the THz generator, a 1-mm thick Teflon filter removed the remaining infrared and visible light. The THz naturally diverged, and then was focused by a 2-inch diameter, 2-inch effective focal length off-axis parabolic mirror to roughly a 500 μm beam radius. At the focus, the electric field was measured using electro-optic sampling with 800-nm probe pulses and a 100-μm (110) GaP crystal bonded to a 1-mm (001) GaP crystal.[24-28] Considering the absorption in the Teflon filter, the active GaP thickness, and the probe pulse duration, the bandwidth of detection is limited to ~6 THz.

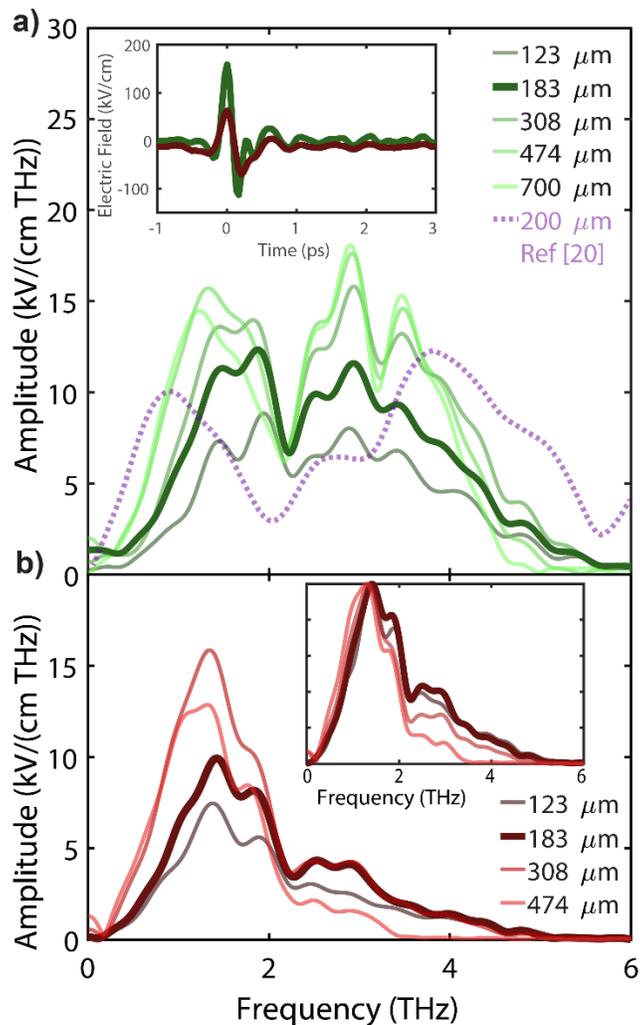

**Fig. 1.** THz spectra generated from BNA crystals of varying thickness using (a) a 1250-nm pump and (b) 800-nm pump. The dashed purple line in (a) is THz generated from a 200-μm thick BNA crystal with a 35 fs, 1150-nm pump, as reported in Ref. 20. The inset to (a) shows time traces for 183 μm crystal pumped with 1250 nm (green) and 800 nm (red). The inset to (b) shows normalized spectra using the 800-nm pump.

Figure 1 shows the THz generation from thin BNA crystals with the specified thicknesses pumped with 1250 nm (Fig. 1(a)) and 800 nm (Fig. 1(b)). The inset to Fig. 1(a) shows the time trace for a 183 μm crystal pumped with 800 nm (red) and 1250 nm (green). There are some important differences in these THz spectra compared to other previous work[19, 20]. For a more direct comparison to recent measurements using a 200-μm thick BNA crystal from Ref. 20, the dotted purple line in Fig. 1(a) (with arbitrary amplitude) shows a spectrum with a large hole centered at ~2 THz. In contrast, the dip in the spectra from our BNA crystals is narrower for the whole range of thicknesses, indicating a significant difference in crystal characteristics. The thinnest of our high-quality crystals generates a very broad and smooth spectrum extending out to 6 THz. We note that the spectrum from Ref. 20 is broader due to a shorter laser pulse duration and higher detection bandwidth, but this does not explain the much larger dip in the spectrum, apparently due to differences in BNA characteristics. Unfortunately, we do not have access to the crystals used in Ref. 20 for a completely comprehensive comparison.

Due to the prevalence of Ti:Sapphire lasers, we also tested the THz output of these thin, high-quality BNA crystals with 800-nm pump light directly from the laser. In Fig. 1(b), we see narrower and more peaked THz spectra, with the maximum spectral amplitude at about 1.5 THz. For the thinnest crystals, the THz frequency content extends out to 5.5 THz, highlighting useful differences compared to previous measurements of thick BNA crystals pumped with 800-nm light, where the spectral amplitude died out at 3 THz[19]. The breadth of each spectrum is clearer in the normalized inset in Fig. 1(b). Thus, even with an 800-nm pump, broadband THz fields can be generated with BNA, showing a useful alternative to tilted-pulse-front LiNbO$_3$ THz sources[8]. With the current limitation due to the damage threshold of BNA (at ~4 mJ/cm$^2$ for 800-nm pump light), the peak field strengths don't yet rival those made with LiNbO$_3$, but the generated field strength exceeds 100 kV/cm and the broader spectrum can prove very valuable for studying a number of samples with relevant excitations in the range from 3-6 THz. The THz spectra shown in Fig. 1 were generated using 8 mJ/cm$^2$ and 1.7 mJ/cm$^2$ fluences for Fig. 1(a) and (b) respectively.

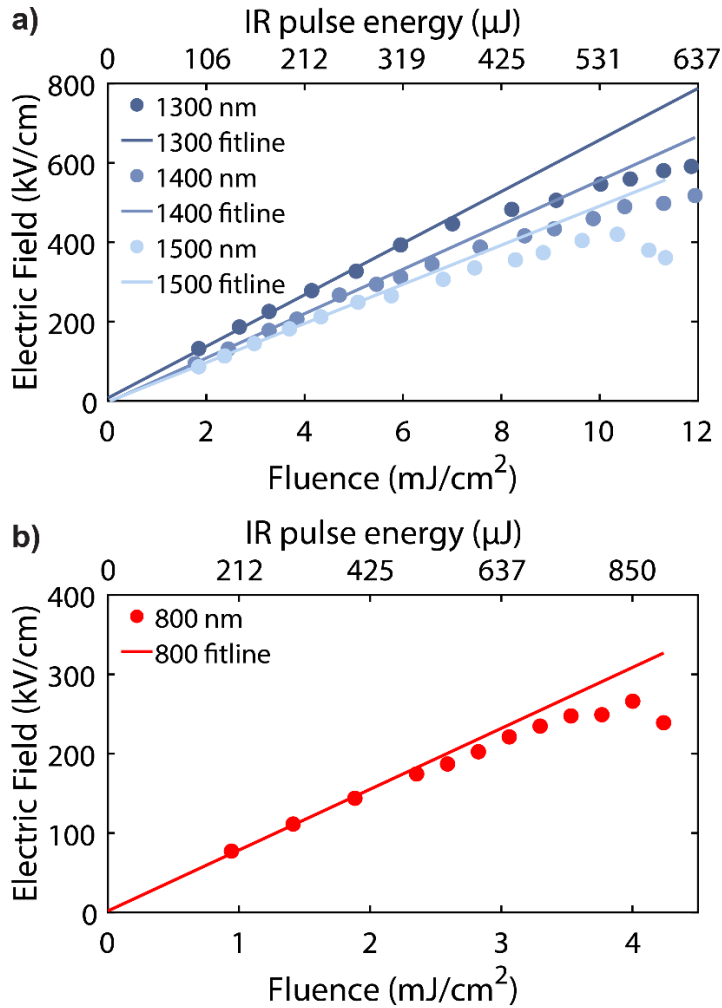

**Fig. 2.** Peak-to-peak electric field as a function of fluence and IR pulse energy. Data collected using EO sampling with GaP. Linear fit-lines were determined using the lowest five fluences for each wavelength and the origin in (a), and the lowest three fluences and the origin in (b).

When using NLO crystals for THz generation, saturation effects can limit the maximum field strengths achievable. Figure 2 shows pump-power dependent measurements of the THz output for different pump wavelengths, over a larger range of fluences than previously reported[20]. We define the beginning of saturation to be the point where there is no longer a linear relationship between pump fluence and output peak electric field. To aid in seeing this, in Fig. 2 we fit a line to the first five points and the origin. For each of the longer wavelengths, saturation effects begin at about 6 mJ/cm$^2$, and for 800 nm, they begin near 2 mJ/cm$^2$, though they are less pronounced.

We note that the generated THz increases moving from 1500 nm down to 1300 nm. For the 1500-nm measurements, the two highest fluences tested were enough to damage the crystal, which is why the last two measurements at this wavelength show a decrease in electric field strength. A similar damage effect is seen with the highest 800-nm fluence of about 4.2 mJ/cm$^2$.

To better understand the generated THz spectra, and for a comprehensive comparison to previous work on BNA, we extracted the THz and optical properties of our high-quality, thin crystals. We extracted the complex refractive index of BNA from 0.5-5.5 THz using THz time-domain spectroscopy. Figure 3 shows the real part of the (a) refractive index and (b) absorption coefficients (dark blue lines), along with Lorentz oscillator fits (light-blue dashed lines). The solid green line in Fig. 3(b) shows the NIR absorption coefficient, with an average value of 4.3 cm$^{-1}$ from 600 to 1800 nm (note the upper x-axis indicating wavelength).

Figure 3(a) offers a comparison with previously reported values of the THz refractive index and the NIR group index of BNA.[22] The filled circles with error bars indicate the group index we measure with an autocorrelator technique[5], and the dotted light-green line shows the group index determined from fits to the data in Ref. 23. A comparison between NIR group index and THz refractive index predicts the observed increase in THz generation moving from 1500-nm down to 1250-nm pump wavelengths because the 1250-nm group index offers better phase matching, particularly in the range from 2.5 to 5.5 THz (see Fig. 3(a)). An extrapolation of the NIR group index data and comparison with our refractive index data suggests that near ideal phase matching for THz generation would be possible with increasingly common 1-micron ultrafast laser sources.

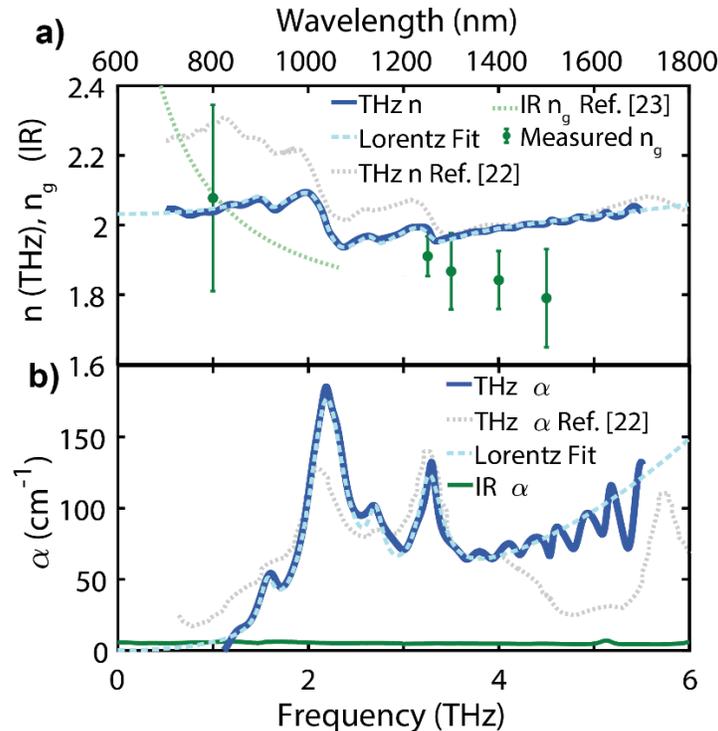

**Fig. 3.** (a) The THz refractive index (solid dark-blue line) and NIR group index (filled green circles). The lighter dashed line indicates a Lorentz-oscillator fit to the refractive index. The dotted green line shows the group index as a function of wavelength from Ref. 23. (b) Absorption coefficient of BNA. The solid dark blue line indicates the THz absorption with the lighter dashed line indicating the Lorentz-oscillator fit. The solid green line shows the NIR absorption coefficient as a function of wavelength, with an average value of 4.3 cm$^{-1}$. The light grey dotted lines in (a) and (b) show THz data from Ref. 22.

Figure 3(a) also shows that 800-nm light has excellent phase-matching characteristics below 2 THz, which leads to the peak in generated THz spectra seen in Fig. 1(b). Because the coherence length scales as $1/\nu$, where $\nu$ is the frequency of the THz light, the worsened phase-matching characteristics of 1250 to 1500 nm light with 0.5-2 THz does not have as much of an effect as the mismatch at higher frequencies does for 800-nm pump light. Therefore, only thin crystals (< 200 μm thick) will produce significant amounts of THz radiation from ~3-6 THz when pumped with 800-nm light.

Figure 3(b) shows the largest absorption coefficient of BNA at ~2.1 THz, which leads to the dip in THz generation at that frequency. Over the entire range, the THz absorption is less than 200 cm$^{-1}$, and at no other point from 0.5-5.5 THz does BNA have an absorption coefficient greater than 150 cm$^{-1}$; this is one reason thin BNA produces such a broad, smooth spectrum. This correspondingly leads to only small changes in refractive index, rising to a maximum of about 2.1 at 2 THz and going to a minimum of about 1.93 at 2.4 THz. The absorption coefficient at NIR wavelengths is relatively low, as seen at the bottom of Fig. 3(b), with an average value of 4.3 cm$^{-1}$ from 600 to 1800 nm.

The THz refractive index and absorption coefficient values we report here differ from what has been previously reported in Ref. 22 (see the dotted grey lines in Fig. 3). However, neither of the previously reported THz refractive indices of (001) oriented BNA reflect the THz generation spectra we observe here. To further validate our refractive index and absorption coefficient values, we modeled THz generation for a variety of thicknesses, as shown in Fig. 4. These modeled spectra match well the measured spectra from Fig. 1(a). Our new refractive index data also explain the large observed spectral differences when pumping BNA with 800-nm light compared to longer wavelengths. We also clearly see the effects of the absorption feature at 2.1 THz, the smaller absorption feature at 3.3 THz, and an absence of any other significant absorption features. This all suggests that, understandably, differences in BNA characteristics will lead to differences in THz generation.

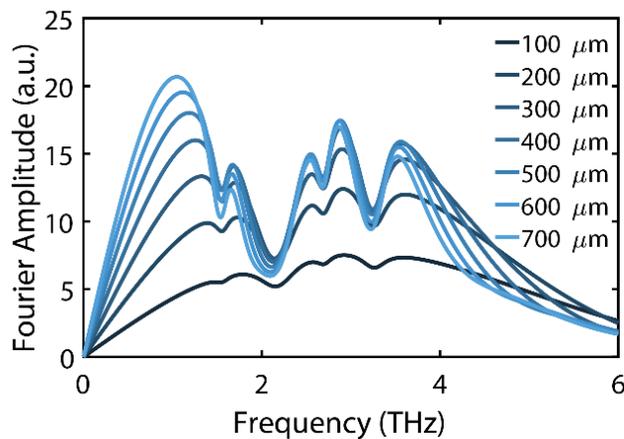

**Fig. 4.** Modeled THz spectra using the 1250-nm group index and the Lorentz oscillator fit to the complex THz refractive index.

Finally, we comment on the recently reported peak electric-field strength of 8 MV/cm for BNA[20]. In that report, Kerr measurements in diamond were used to determine this unusually high electric-field strength produced from a mid-range amplified laser system. THz Kerr measurements in diamond have been used to analyze THz waveforms in a variety of reported measurements[13, 20, 29, 30], and we decided to compare using diamond to calibrate the peak THz electric field with common electro-optic detection. We measured the THz electric field with GaP as described above, and then measured again using 250-μm thick diamond for Kerr effect measurements. The measured probe phase shift in diamond ($\Delta\phi$) is shown in Fig. 5(a). To extract the electric field from the diamond THz Kerr measurements (shown on the right axis), we used the Kerr coefficient from Ref. 29 of $3 \times 10^{16}$ cm$^2$/W (note that there is a typo of this value in Ref. 30). The standard EO-sampling measurement yielded a peak electric field of 300 kV/cm, while the diamond measurement yielded a peak electric field of 1.6 MV/cm, about 5 times greater than what EO-sampling indicates. We recognize that there may be differences in peak electric fields determined by electro-optic sampling (see Ref. 29 for a nice brief overview), but often this is on the order of 10–30%, and therefore a 500% difference in peak field strength is worth further consideration.

The Kerr effect signal is proportional to $E(t)^2$. Presumably, a larger recovered field in diamond could result due to better phase-matching between the THz and 800-nm probe light, leading to a better measurement of higher frequency components. The inset to Fig. 5(b) shows the normalized diamond Kerr trace together with $E(t)^2$ calculated from the EO measurements. The main panel of Fig. 5(b) shows the Fourier transform of $E(t)^2$ and the Kerr signal from diamond to see if diamond captures any higher-frequency components. The diamond frequency response drops much more quickly than the GaP response, indicating that GaP more accurately recovers higher THz frequency components; the general structure of the two signals is comparable, but there are, however, much higher frequency components present in the EO-signal that appear to be lost in the diamond Kerr effect measurement. Thus, such a large discrepancy in peak fields is likely not due to the diamond Kerr measurements capturing higher field-strength components that are absent in the 100-μm GaP EO sampling measurements.

We note that the measurements in Ref. 29 emphasize that the diamond Kerr constant is for 1 THz, and in that work, the bandwidth of the diamond measurements is sufficient to capture the lower bandwidth of LiNbO$_3$ generated THz pulses. A careful examination of additional published diamond measurements at higher THz frequencies reveal similarities to what we see here[31, 32], likely indicating that there is some unexplored frequency dependence to $\chi^{(3)}$. This all suggests that until a more complete characterization of THz diamond Kerr effect measurements can be carried out, we recommend caution in the calibration of peak electric fields based on this method.

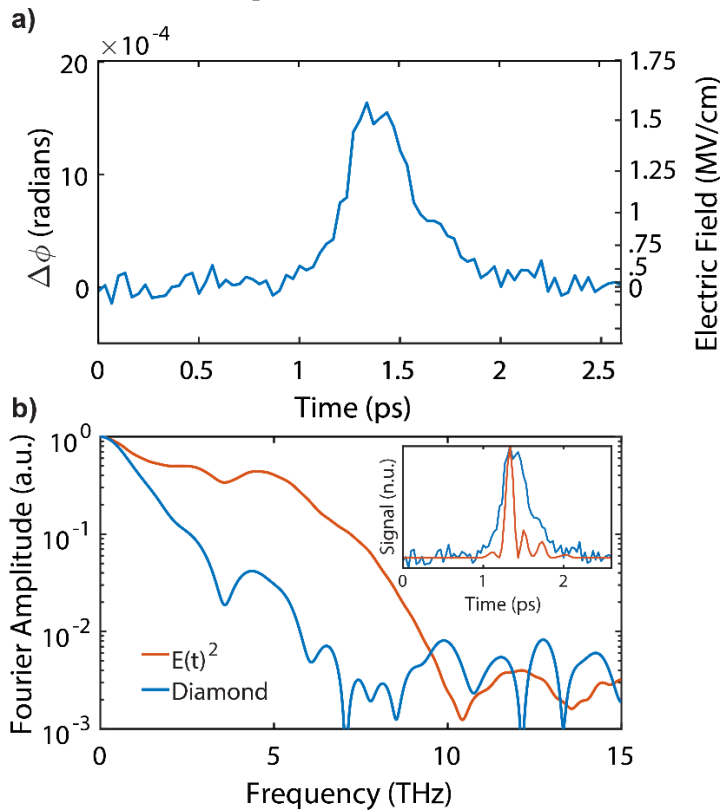

**Fig. 5.** (a) THz Kerr effect signal in diamond produced from BNA pumped with 1250-nm light. The right axis shows the potentially overstated electric field based on this measurement. In (b), the Fourier transform of the diamond measurement and the square of the electric field as measured with GaP are shown (both normalized). Inset to (b): normalized diamond Kerr signal compared to $E(t)^2$ from EO-sampling.

In conclusion, BNA is an excellent option for creating a smooth, broadband spectrum for THz spectroscopy. With no absorption features above 200 cm$^{-1}$ and good phase-matching with 1250-nm NIR light, BNA crystals can produce significant amounts of THz between 0-6 THz. However, as we compare THz generation and optical properties of our THz crystals with other published data, we see that differences in BNA crystals can result in less optimal spectral properties. Furthermore, thin, high-quality BNA crystals pumped with the 800-nm output of common Ti:Sapphire amplifiers can also produce intense, broadband THz pulses. While it doesn't yet rival the peak-field strengths obtained by tilted-pulse front pumping of LiNbO$_3$, thin BNA crystals provide an alternative broader bandwidth and still relatively high-field alternative to LiNbO$_3$-based THz sources.

†Contributed equally to this work.

**Acknowledgements.** We thank the Department of Chemistry and Biochemistry at Brigham Young University for funding. We also express appreciation to Tobias Kampfrath and Steven Johnson for useful discussions.

**Disclosures.** D. J. Michaelis and J. A. Johnson are co-founders of Terahertz Innovations, LLC.